# Existence of Invariant Tori for Differentiable Hamiltonian Vector Fields without Action-Angle Variables

Jong, Wu Hwan[*] and Paek, Jin Chol[**]

\* Information Technology Institute, **KIM IL SUNG** University, DPRK
\*\* Faculty of Mathematics, **KIM IL SUNG** University, DPRK

**Abstract**: We proved the theorem for existence of invariant tori in differentiable Hamiltonian vector fields without action-angle variables. It is a generalization of the result of [Llave, 2005] that deals with analytic vector fields.

## 1. Introduction

In this paper, we provide a KAM theorem on existence of invariant tori with a Diophantine vector for differentiable Hamiltonian vector fields. We studied differentiable Hamiltonian vector fields which may not be perturbations of integrable vector fields or may not be written in action-angle variables.

The existence problem of invariant tori for Hamiltonian system is often appeared in the study of stability problem of system of planets in celestial mechanics ([Celletti-Chierchia 2006], [Fèjoz 2004], [Locatelli-Giorgilli 2007]), the calculation of long time stability region in particle accelerator ([Helleman, 1985]), the analysis of relaxation times in analysis of relaxation phenomena in molecular dynamics ([Casartelli 1983]) and the analysis of the complicated orbit structure exhibited by several classes of population models ([Gidea, 2009]) and so on.

[Kolmogorov 1954] has proposed the procedure which clarifies the existence of invariant tori for perturbed analytical Hamiltonian vector field with action-angle variables at first and [Arnold 1963] has given the rigorous proof based on Kolmogorov's procedure. [Moser 1961, 1962] has proved the existence of invariant tori for analytical area-preserving twist mappings on 2-dimensional annulus with action-angle variables and moreover he has relaxed the assumption of analyticity of the map to $C^{333}$ differentiability. After that, [Rüssmann 1970] has relaxed the differentiability condition with the existence of invariant tori in Hamiltonian system to $C^5$ class and [Takens 1971] has clarified that it is not enough for $C^1$ class. Finally [Herman 1986] has clarified that it is enough for $C^3$ class but not to $C^2$-mappings whose second derivatives belong to the Hölder class $C^{1-\delta}$ where $\delta > 0$ is small. However these are all the results for the Hamiltonian systems which are perturbations of some integrable systems and written in action-angle variables. Readers can refer to good expository article [Llave 2001] for KAM theory.

On the other hands, action-angle variables have singularity at elliptic fixed points or in neighborhood of separatrix and the use of action-angle variables are too restrictive in the case of numerical analysis. Moreover, in many practical applications, we have to consider the system which is not near to integrable one but has approximate invariant tori with sufficiently small error ([Llave, 2005]).

[Llave, 2005] neither proved the existence of invariant tori for *analytic* Hamiltonian systems which are neither perturbed integrable systems nor written in action-angle variables. [Haro-Llave 2004] applied this result in the numerical computation of invariant tori.

[Llave et al 2008] considered finitely *differentiable* symplectic *maps* without action-angle variables.

In this paper, we prove that there exists true invariant tori nearby approximately invariant tori for *differentiable* Hamiltonian *vector fields* which are neither perturbed integrable vector fields nor written in action-angle variables.

## 2. Some definitions and the existence of invariant tori in the case of analytic Hamiltonian vector field

Given $\gamma > 0$ and $\sigma > n-1$, we define $D_n(\gamma, \sigma)$ as the set of frequency vectors $\omega = (\omega_1, \cdots, \omega_n) \in \mathbf{R}^n$ satisfying the *Diophantine condition*:

$$|k \cdot \omega| \geq \gamma |k|^{-\sigma}, \ \forall k \in \mathbf{Z}^n - \{0\}$$

where $|k| = |k_1| + \cdots + |k_n|$. Let $U_\rho$ denote the complex closed strip of width $\rho > 0$, that is, $U_\rho = \{\theta \in \mathbf{C}^n \ ; \ |\mathrm{Im}\,\theta| \leq \rho\}$. Let $\mathcal{B}$ is a compact subset of $\mathbf{R}^n$ which is included in closure of its interior. Given a function $g \in C^m(\mathcal{B})$, for $m \in \mathbf{Z}_+ = \mathbf{N} \cup \{0\}$ we will denote the $C^m$-norm of $g$ on $\mathcal{B}$ by $|g|_{C^m, \mathcal{B}}$. Given a 1-periodic function $K$, continuous on $\mathbf{T}^n = \mathbf{R}^n / \mathbf{Z}^n$, we denote the average of $K$ on $\mathbf{T}^n$ by

$$<K> = \int_{\mathbf{T}^n} K(\theta) d\theta.$$

If $X$ is a open subset of topological space of $M$ then we denote this fact by $X \overset{\circ}{\subset} M$. We will assume that $\mathbf{U}$ is either $\mathbf{T}^n \times U$ with $U \overset{\circ}{\subset} \mathbf{R}^n$ or $\mathbf{U} \overset{\circ}{\subset} \mathbf{R}^{2n}$. The results for a Hamiltonian vector field with differential Hamiltonian function $H: \mathbf{U} \to \mathbf{R}$ are based on the study of the equation

$$\partial_\omega K(\theta) = J \nabla H(K(\theta)), \tag{1}$$

where $J = \begin{pmatrix} 0 & I_n \\ -I_n & 0 \end{pmatrix}$, $I_n$ is $n$-dimensional unit matrix and $\partial_\omega$ is the derivative in direction $\omega$:

$$\partial_\omega K = \sum_{i=1}^n \omega_i \frac{\partial K}{\partial \theta_i}.$$

In the above $K: \mathbf{T}^n \to \mathbf{U}$ is the function to be determined and $\omega \in D(\gamma, \sigma) \subset \mathbf{R}^n$.

Note that (1) implies that the $K(\mathbf{T}^n)$ is invariant under the Hamiltonian vector field with Hamiltonian function $H$, $X_H = J \nabla H$.

**Definition 1.** $\mathcal{P}_\rho$ will denote the Banach space set of functions $K: U_\rho \to \mathbf{U}$ which are 1-periodic in all its variables, real analytic on the interior of $U_\rho$, continuous on the boundary of $U_\rho$ with norm $\|K\|_\rho = \sup_{\theta \in U_\rho} |K(\theta)|$ where $|\cdot|$ represents the maximum norm on the space $\mathbf{C}^m$.

**Definition 2.** $K \in \mathcal{P}_\rho$ is said to be *non-degenerate* if it satisfies the following two conditions:

1) There exists a $n \times n$ matrix-valued function $N(\theta)$ satisfying
$$N(\theta)DK(\theta)^T DK(\theta) = I_n.$$

2) The average of $S$, $<S>$ is invertible with
$$S(\theta) = N(\theta)DK(\theta)^T[A(\theta)J - JA(\theta)]DK(\theta)N(\theta)$$

where
$$A(\theta) = DX_H(K(\theta)) = \begin{pmatrix} D_x \nabla_y H(K(\theta)) & D_y \nabla_y H(K(\theta)) \\ -D_x \nabla_x H(K(\theta)) & -D_y \nabla_x H(K(\theta)) \end{pmatrix}.$$

**Theorem** (theorem 5 in [Llave et al. 2005]). Let $\omega \in D(\gamma, \sigma)$. Assume that $K_0 \in \mathcal{P}_\rho$ is non-degenerate. Assume that $H$ is real analytic and that is can be holomorphically extended to some neighborhood of the image of $U_\rho$ under $K_0$:
$$\mathcal{B}_r = \mathcal{B}_r(K_0) := \{z \in \mathbf{R}^n \; ; \; \inf_{\theta \in U_\rho} |z - K_0(\theta)| < r\}.$$

Define the error function by
$$e_0(\theta) = J\nabla H(K_0(\theta)) - \partial_\omega K_0(\theta).$$

There exists a constant $c > 0$, depending on $\sigma$, $n$, $|H|_{C^3, \mathcal{B}_r}$, $\|DK_0\|_\rho$, $\|N_0\|_\rho$, $|<S_0>^{-1}|$, such that if
$$c\gamma^{-4}\delta_0^{-4\sigma} \|e_0\|_\rho < 1 \qquad (2)$$

and
$$c\gamma^{-2}\delta_0^{-2\sigma} \|e_0\|_\rho < r \qquad (3)$$

with $\delta_0 = \min(1, \rho/12)$, then there exists a solution for (1), $K_\infty$, which is real analytic on $U_{\rho/2}$ and satisfies the non-degenerate conditions. Moreover
$$\|K_\infty - K_0\|_{\rho/2} \leq r \qquad (4)$$

and $\|DK_\infty\|_{\rho/2}$, $\|N_\infty\|_{\rho/2}$, $|<S_\infty>^{-1}|$ satisfies the following inequalities
$$\|DK_\infty\|_{\rho/2} \leq \|DK_0\|_{\rho_0} + \beta$$
$$\|N_\infty\|_{\rho/2} \leq \|N_0\|_{\rho_0} + \beta$$
$$|<S_\infty>^{-1}| \leq |<S_0>^{-1}|_{\rho_0} + \beta.$$

where $\|N_\infty\|_{\rho/2}$, $|<S_\infty>^{-1}|$ is given by definition 2, replacing $K$ with $K_\infty$ and $\beta = \gamma^{-2}\delta_0^{2\sigma-1}2^{-4\sigma}$.

**Remark 1.** The dependence of constant $c$ on $|H|_{C^3, \mathcal{B}_r}$, $\|DK_0\|_\rho$, $\|N_0\|_\rho$ and $|<S_0>^{-1}|$ is polynomial. That is, there exists a polynomial, $\lambda(y_1, y_2, y_3, y_4)$ with positive coefficients depending on $\sigma$, $n$ and such that
$$c = \lambda(|H|_{C^3, \mathcal{B}_r}, \|DK_0\|_\rho, \|N_0\|_\rho, <S_0>^{-1}|).$$

(remark 15 in [Llave et al. 2005]).

## 3. The existence of invariant tori in the case of Differentiable Hamiltonian vector field

To prove the existence of invariant tori in the case of differentiable Hamiltonian, we use some approximation theorems.

**Lemma 1.** Let $l \in \mathbf{N}$, $I = [0,1]$ and $f \in C^{l+1}([0, 1])$, $(l \geq 3)$. Then there exists a sequence of analytic function $\{f_k\}$ converges to $f$ on $I$ such that

$$|f_k - f|_{C^l, [0, 1]} \leq Ck^{-1}(|f|_{C^{l+1}} + 1),$$

where $c$ is a constant depended on only $l$.

**Proof**. See ([Lorentz 1986]).

**Lemma 2.** The $q$-th order derivatives of the Bernstein polynomials

$$f_k(x) = \sum_{p=0}^{k} C_k^p f(\frac{p}{k}) x^p (1-x)^{k-p}$$

of function $f \in C^l(I)$ are as follows

$$f_k^{(q)}(x) = k(k-1)\cdots(k-q+1) \sum_{p=0}^{k-q} \Delta^q f(\frac{p}{k}) C_{k-q}^p x^p (1-x)^{k-p-q}, \quad (q = 1, 2, \cdots, k),$$

where $\Delta^q f(x) = f(x + \frac{q}{k}) - C_q^1 f(x + \frac{q-1}{k}) + \cdots + (-1)^q f(x)$.

**Proof.** See [Lorentz 1986].

**Lemma 3.** Let $B_n = [a_1, b_1] \times [a_2, b_2] \times \cdots \times [a_n, b_n]$ and $f \in C^4(B_n)$. Then there exists a sequence of analytic function $\{f_k\}$ on $B_n$ such that $|f_k - f|_{C^3, B_n} \to 0, \; k \to \infty$.

**Proof**. First, we will prove that it is sufficient to prove the lemma for

$$E_n = \underbrace{[0, 1] \times [0, 1] \times \cdots \times [0, 1]}_{n}.$$

In fact, the linear transformation

$$\begin{cases} y_1 = \dfrac{x_1 - a_1}{b_1 - a_1} \\ \quad \vdots \\ y_n = \dfrac{x_n - a_n}{b_n - a_n} \end{cases} \quad (5)$$

transforms the function $f : (x_1, \cdots, x_n) \in B_n \to f(x_1, \cdots, x_n) \in \mathbf{R}$ into $\bar{f} : (y_1, \cdots, y_n) \in E_n \to f(y_1, \cdots, y_n) \in \mathbf{R}$. Let's sequence of analytic functions $\{\bar{f}_k\}$ converges to $\bar{f}$ in the sense of $C^3$-norm. Then by applying the inverse transformation of (5) to $\bar{f}_k$, we obtain the sequence of functions

$$f_k(x_1, \cdots, x_n) = \bar{f}_k(\frac{x_1 - a_1}{b_1 - a_1}, \cdots, \frac{x_n - a_n}{b_n - a_n})$$

analytic in $B_n$ and converges to $f$ in the sense of $C^3$-norm. After all, it is sufficient to prove the lemma for $E_n$.

We prove the lemma, using induction. If $n = 1$, then from the lemma 1 the Bernstein polynomials of $f$, $f_k(x) = \sum_{p=0}^{k} C_k^p f(\frac{p}{k}) x^p (1-x)^{k-p}$ satisfies the statement of lemma 3. Now assume that the lemma holds for $n$, then we will prove the lemma for $n+1$. For fixed point $(x_1, \cdots, x_n) \in E_n = [0, 1] \times \cdots \times [0, 1]$, let us consider the Bernstein polynomial of $f$ with respect to only $x_{n+1}$

$$g_k(x_1, \cdots, x_n, x_{n+1}) = \sum_{p=0}^{k} C_k^p f(x_1, \cdots, x_n, \frac{p}{k}) x_{n+1}^p (1-x_{n+1})^{k-p}.$$

Then from the lemma 1, there exists a const $C$ such that for an arbitrarily fixed point $(x_1, \cdots, x_n) \in E_n$, the following inequality holds:

$$|g_k(x_1, \cdots, x_n, \cdot) - f(x_1, \cdots, x_n, \cdot)|_{C^3, [0,1]} \leq C \cdot \left( |f(x_1, \cdots, x_n, \cdot)|_{C^4, [0,1]} + 1 \right) \cdot \frac{1}{k}.$$

Define $M = |f|_{C^4, E_{n+1}} + 1$, then $|f(x_1, \cdots, x_n, \cdot)|_{C^4, [0,1]} + 1 \leq M$ and therefore

$$|g_k(x_1, \cdots, x_n, \cdot) - f(x_1, \cdots, x_n, \cdot)|_{C^3, [0,1]} \leq \frac{CM}{k}.$$

i.e. for an arbitrary point $(x_1, \cdots, x_n) \in E_n$,

$$|g_k(x_1, \cdots, x_n, \cdot) - f(x_1, \cdots, x_n, \cdot)|_{C^3, [0,1]} \leq \varepsilon_k \tag{6}$$

where $\varepsilon_k = \frac{CM}{k}$.

The sequence of Bernstein polynomials of $f$ converges to $f$ depending only the norm of $f$, hence for any $i, j, h = 1, \cdots, n$, the inequalities

$$|D_i g_k(x_1, \cdots, x_n, \cdot) - D_i f(x_1, \cdots, x_n, \cdot)|_{C^2, [0,1]} \leq \frac{C|f|_{C^4, E_{n+1}}}{k}$$

$$|D_i D_j g_k(x_1, \cdots, x_n, \cdot) - D_i D_j f(x_1, \cdots, x_n, \cdot)|_{C^1, [0,1]} \leq \frac{C|f|_{C^4, E_{n+1}}}{k}$$

$$|D_i D_j D_h g_k(x_1, \cdots, x_n, \cdot) - D_i D_j D_h f(x_1, \cdots, x_n, \cdot)|_{C^0, [0,1]} \leq \frac{C|f|_{C^4, E_{n+1}}}{k}$$

hold. In fact, $D_i g_k$ is the Bernstein polynomial of $D_i f$, hence from the lemma 1

$$|D_i g_k(x_1, \cdots, x_n, \cdot) - D_i f(x_1, \cdots, x_n, \cdot)|_{C^2, [0,1]} \leq \frac{C(|Df(x_1, \cdots, x_n, \cdot)|_{C^3, [0,1]} + 1)}{k}$$

$$\leq \frac{C(|Df|_{C^3, E_{n+1}} + 1)}{k}.$$

Other above two inequalities hold similarly.

Define $f_{k,p}(x_1, \cdots, x_n) = f(x_1, \cdots, x_n, \frac{p}{k})$, then $f_{k,p}$ is a $n$-variable function. Hence, by the assumption for the induction, there exists a sequence of functions $\{f_{k,p}^j\}_{j=1}^{\infty}$ analytic on $E_n$ such that $f_{k,p}^j$ converges to $f_{k,p}$ in the $C^3$ norm.

We will define $j \stackrel{def}{=} j(k, p)$ as a least number satisfying the following inequality

$$|f_{k,p}^j - f_{k,p}|_{C^3, E_n} < \frac{\varepsilon_k}{8(k+1)k(k-1)(k-2)}, \quad (k \geq 3) \tag{7}$$

and define

$$f_k(x_1, \cdots, x_n, x_{n+1}) \stackrel{def}{=} \sum_{p=0}^{k} C_k^p f_{k,p}^{j(k,p)}(x_1, \cdots, x_n) x_{n+1}^p (1-x_{n+1})^{k-p}.$$

Then $\{f_k\}_{k=3}^{\infty}$ become a sequence of analytic functions converges to $f$ on $E_{n+1}$ in $C^3$ norm. In fact, the inequality

$$|f_k - f|_{C^0, E_{n+1}} \leq |f_k - g_k|_{C^0, E_{n+1}} + |g_k - f|_{C^0, E_{n+1}} \leq (k+1) \cdot \frac{\varepsilon_k}{8(k+1)k(k-1)(k-2)} + \varepsilon_k \leq 2\varepsilon_k$$

holds, hence $f_k$ converges to $f$ on $E_{n+1}$ uniformly.

Next, for any $i=1, \cdots, n$, we have

$$|D_i f_k(x_1, \cdots, x_n, x_{n+1}) - D_i f(x_1, \cdots, x_n, x_{n+1})| \leq$$
$$\leq |D_i f_k(x_1, \cdots, x_n, x_{n+1}) - D_i g_k(x_1, \cdots, x_n, x_{n+1})| +$$
$$+ |D_i g_k(x_1, \cdots, x_n, x_{n+1}) - D_i f(x_1, \cdots, x_n, x_{n+1})| \leq$$
$$\leq \sum_{p=0}^{k} C_k^p |D_i f_{k,p}^{j(k,p)}(x_1, \cdots, x_n) - D_i f_{k,p}(x_1, \cdots, x_n)| x_{n+1}^p (1-x_{n+1})^{1-p} + \varepsilon_k \leq$$
$$\leq \sum_{p=0}^{k} C_k^p |f_{k,p}^{j(k,p)} - f_{k,p}|_{C^1, E_n} x_{n+1}^p (1-x_{n+1})^{1-p} + \varepsilon_k \leq$$
$$\leq (k+1) \cdot \frac{\varepsilon_k}{8(k+1)k(k-1)(k-2)} + \varepsilon_k \leq 2\varepsilon_k.$$

Hence for $i=1, \cdots, n$ $D_i f_k$ $C^0$-converges to $D_i f$ on $E_{n+1}$.

From lemma 2, the derivative of $g_k$ on $n+1$-variable is denoted by

$$D_{n+1}^q g_k = k(k-1)\cdots(k-q+1)\sum_{p=0}^{k-q} \Delta^q f(x_1, \cdots, x_n, \frac{p}{k}) C_{k-q}^p x_{n+1}^p (1-x_{n+1})^{k-q-p}$$

Where

$$\Delta^q f(x) = f(x_1, \cdots, x_n, x_{n+1} + \frac{q}{k}) - C_k^1 f(x_1, \cdots, x_n, x_{n+1} + \frac{q-1}{k}) + \cdots + (-1)^k f(x).$$

From this expression, the following inequality holds:

$$|D_{n+1}^3 f_k(x_1, \cdots, x_n, x_{n+1}) - D_{n+1}^3 f(x_1, \cdots, x_n, x_{n+1})|_{C^0, E_{n+1}} \leq$$
$$\leq |D_{n+1}^3 f_k(x_1, \cdots, x_n, x_{n+1}) - D_{n+1}^3 g_k(x_1, \cdots, x_n, x_{n+1})|_{C^0, E_{n+1}} +$$
$$+ |D_{n+1}^3 g_k(x_1, \cdots, x_n, x_{n+1}) - D_{n+1}^3 f(x_1, \cdots, x_n, x_{n+1})|_{C^0, E_{n+1}} \leq$$
$$\leq k(k-1)(k-2) \sum_{p=0}^{k-3} \left[ \left| \Delta^3 f_{k,p}^{l(k,p)}\left(x_1,\cdots,x_n,\frac{p}{k}\right) - \Delta^3 f_{k,p}\left(x_1,\cdots,x_n,\frac{p}{k}\right) \right| \cdot C_{k-3}^p x^p (1-x)^{k-3-p} \right] + \varepsilon_k \leq$$
$$\leq 8k(k-1)(k-2)(k-2) \cdot \frac{\varepsilon_k}{8(k+1)k(k-1)(k-2)} + \varepsilon_k \leq 2\varepsilon_k.$$

Therefore $D_{n+1}^3 f_k$ $C^0$-converges to $D_{n+1}^3 f$ on $E_{n+1}$.

Similarly $D_{n+1}^2 f_k$, $D_{n+1}^1 f_k$ also respectively $C^0$-converges to $D_{n+1}^2 f$, $D_{n+1}^1 f$ on $E_{n+1}$.

Next for any $i, j = 1, \cdots, n$

$$|D_i D_j D_{n+1} f_k(x_1, \cdots, x_n, x_{n+1}) - D_i D_j D_{n+1} f(x_1, \cdots, x_n, x_{n+1})|_{C^0, E_{n+1}} \leq$$
$$\leq |D_i D_j D_{n+1} f_k(x_1, \cdots, x_n, x_{n+1}) - D_i D_j D_{n+1} g_k(x_1, \cdots, x_n, x_{n+1})|_{C^0, E_{n+1}} +$$
$$+ |D_i D_j D_{n+1} g_k(x_1, \cdots, x_n, x_{n+1}) - D_i D_j D_{n+1} f(x_1, \cdots, x_n, x_{n+1})|_{C^0, E_{n+1}} \leq$$
$$\leq k \cdot k \sum_{p=0}^{k-1} |\Delta^1 D_i D_j f_{k,p}^{l(k,p)}(x_1,\cdots,x_n,\frac{p}{k}) - \Delta^1 D_i D_j f_{k,p}(x_1,\cdots,x_n,\frac{p}{k})| \cdot C_{k-1}^p x^p (1-x)^{k-1-p} + \varepsilon_k \leq$$
$$\leq 2k \cdot k \cdot \frac{\varepsilon_k}{8(k+1)k(k-1)(k-2)} + \varepsilon_k \leq 2\varepsilon_k.$$

hence $D_i D_j D_{n+1} f_k$, $D_i D_{n+1}^2 f_k$, $D_i D_{n+1} f_k$ also respectively converges to $D_i D_j D_{n+1} f$, $D_i D_{n+1}^2 f$, $D_i D_{n+1} f$ on $E_{n+1}$ uniformly. After all, all the 3rd derivatives of $f_k$ $C^0$-converges to $f$ on $E_{n+1}$. i.e.

$$|f_k - f|_{C^3, D_n} \to 0, \ k \to \infty. \ \square$$

**Remark 2.** [Moser 1966] proved the existence of sequence of analytic functions which converges to the given $2\pi$-periodic $C^l$ function using analytic approximation by trigonometric polynomials for the function in Lemma 1 in section 7 in chapter 3. [Zehnder 75] proved the existence of sequence of analytic functions which converges to the given continuous function in $C^0$ norm in Lemma 2.1 at pp.110. [Salamon 2004] gave analytic smoothing theorem in lemma 3 in section 3. However, this result required that the function is defined in $\mathbf{R}^n$. Through the proof, they use the Fourier Transform, so this condition is essential. In our paper, there is not the assumption for periodicity of Hamiltonian and the Hamiltonian is not defined on $\mathbf{R}^n$ generally. And we had to prove the existence of sequence of analytic functions which converges to $C^l$ function($l \geq 4$) in $C^3$ norm in the process of theory development. The lemma 3 is just the proposition proving this fact.

**Lemma 4.** Let $l \in \mathbf{N}$. If a sequence of functions $\{f_k(x)\}$ analytic on $U_{\rho/4^{k-1}}$ satisfies the following inequality

$$\|f_k(x) - f_{k+1}(x)\|_{\rho/4^k} \leq A(4^{-l})^k,$$

where $A \geq 0$ is a suitable constant, then $f_k(x)$ converges to certain function $f \in C^l(\mathbf{T}^n)$.

**Proof.** See lemma 1 of Chapter 3, Section 7 in [Moser 1966].

**Theorem 1.** Let $\omega \in D(\gamma, \sigma)$ ($\sigma > n-1$). Assume that $K_0 \in \mathcal{P}_\rho$ is analytic in int$U_\rho$ and non-degenerate. Let $U' \subset \mathbf{R}^n$ and $U \subset U'$ be open sets and $r > 0$. We put $\mathbf{U}' = \mathbf{T}^n \times U'$ and $\mathbf{U} = \mathbf{T}^n \times U$. We assume $B_{3r}(U) \subset \mathbf{U}'$ and $H : \mathbf{U}' \to \mathbf{R}$ is $C^l$ ($l \geq 4$). Define the error function

$$e_0(\theta) \stackrel{def}{=} J\nabla H(K_0(\theta)) - \partial_\omega K_0(\theta).$$

Then there exists a constant $c > 0$, depending on $\sigma$, $n$, $|H|_{C^3, \mathcal{B}_r}$, $\|DK_0\|_\rho$, $\|N_0\|_\rho$, $|<S_0>^{-1}|$, such that if

(2): $c\gamma^{-4}\delta_0^{-4\sigma} \|e_0\|_\rho < 1$

and

(3): $c\gamma^{-2}\delta_0^{-2\sigma} \|e_0\|_\rho < r$

with $\delta_0 = \min(1, \rho/12)$, then there exists a solution for (1), $K_\infty : \mathbf{T}^n \to \mathbf{U}$, which is $C^1$ on $\mathbf{T}^n$ and satisfies the non-degenerate conditions.

**Proof.** We use the following notation.

$$\mu_0 = |H|_{C^3, \mathcal{B}_{2r}}, \ d_0 = \|DK_0\|_\rho, \ v_0 = \|N_0\|_\rho, \ \tau_0 = |<S_0>^{-1}|$$

$$\beta = \gamma^{-2}\delta_0^{2\sigma-1}\frac{1}{2^{4\sigma} - 2^{2\sigma+1}}, \ \mu = \mu_0 + 1, \ d = d_0 + \beta, \ v = v_0 + \beta, \ \tau = \tau_0 + \beta + 1$$

and define

$$c \stackrel{def}{=} \lambda(\mu, d, v, \tau).$$

$\mathcal{B}_{3r}(K_0)$ is a bounded subset of $\mathbf{R}^n$, thus there exists a rectangle region $B_{2n}(K_0) = [a_1, b_1] \times \cdots \times [a_{2n}, b_{2n}]$ satisfying the following relation

$$\mathcal{B}_{3r}(K_0) \subset B_{2n}(K_0).$$

And $\mathcal{B}_{\frac{5r}{2}}(K_0)$ is a open neighbor of $K_0(U_\rho)$, hence there exists a $\varphi \in C^\infty(B_{2n}(K_0))$ such that

$$\varphi(z) = \begin{cases} 1, & z \in K_0(U_\rho) \\ 0, & z \in B_{2n}(K_0) \setminus \mathcal{B}_{\frac{5r}{2}}(K_0) \end{cases}$$

$$0 < \varphi(z) < 1, \quad z \in B_{2n}(K_0).$$

Now we extend the function $H$ onto $B_{2n}(K_0)$ arbitrarily and consider the function $H \cdot \varphi : B_{2n}(K_0) \to \mathbf{R}$. Because $H$, $\varphi$ are both $C^l$ and $\varphi(z) = 0, z \in B_{2n}(K_0) \setminus \mathcal{B}_{\frac{5r}{2}}(K_0)$, therefore $H \cdot \varphi$ is $C^l$ on $B_{2n}(K_0) \setminus \mathcal{B}_{\frac{5}{2}r}(K_0)$. Thus $H \cdot \varphi$ is $C^l$ on $B_{2n}(K_0)$. $H \cdot \varphi$ preserves function values of $H$ on $\mathcal{B}_{2r}(K_0)$. So we denote $H \cdot \varphi$ as $H$. Then from the lemma 3, there exists a sequence of functions analytic in $B_{2n}(K_0)$ such that

$$|H_k - H_{k+1}|_{C^3, B_{2n}(K_0)} \le A(4^{-k})^{l+2\sigma}$$

$$|H_k - H|_{C^3, B_{2n}(K_0)} \le A(4^{-k})^{l+2\sigma}$$

where $A \ge 0$ is a constant depending on only $|H|_{C^3, B_{2n}(K_0)}$.

And $\mathcal{B}_{2r}(K_0) \subset B_{2n}(K_0)$, thus the following inequalities hold

$$|H_k - H_{k+1}|_{C^3, \mathcal{B}_{2r}(K_0)} \le A(4^{-k})^{l+2\sigma}$$

$$|H_k - H|_{C^3, \mathcal{B}_{2r}(K_0)} \le A(4^{-k})^{l+2\sigma}.$$

Let us prove the theorem in two steps.

**First Step.**

For the first step, we will prove that for some number $k_0$, there exists a solution $K_{k_0}$ of (1) with Hamiltonian $H_{k_0}$ satisfying

$$\|K_{k_0} - K_0\|_{\rho/2} \le r.$$

For this, let us prove that a pair $H_{k_0}$ and $K_0$ satisfy the assumption of proposition 1, i.e. $c_{k_0}$ and $e_{k_0}$ defined by

$$c_{k_0} = \lambda(|H_{k_0}|_{C^3, \mathcal{B}_{2r}(K_0)}, \|DK_0\|_\rho, \|N_0\|_\rho, |<S_0^{k_0}>^{-1}|)$$

$$e_{k_0} = J\nabla H_{k_0}(K_0(\theta)) - \partial_\omega K_0(\theta),$$

satisfy (2) and (3). First let us prove $c_{k_0} < c$. Performing some simple computations, we obtain

$$S_0^{k_0}(\theta) = N_0(\theta)DK_0(\theta)^T [A_{k_0}(\theta)J - JA_{k_0}(\theta)]DK_0(\theta)N_0(\theta) =$$
$$= N_0(\theta)DK_0(\theta)^T [(A_0(\theta)J - JA_0(\theta)) + (A_{k_0}(\theta) - A_0(\theta))J - J(A_{k_0}(\theta) - A_0(\theta))],$$

$$DK_0(\theta)N_0(\theta) =$$
$$= S_0(\theta) + N_0(\theta)DK_0(\theta)^T [(A_{k_0}(\theta) - A_0(\theta))J - J(A_{k_0}(\theta) - A_0(\theta))]DK_0(\theta)N_0(\theta) =$$
$$= S_0(\theta) + \Phi_0(\theta).$$

From the $\|\Phi_0(\theta)\|_{\rho_0} \le 2d^2\nu^2 |H_{k_0} - H|_{C^3, \mathcal{B}_{2r}(K_0)}$ and $|H_{k_0} - H|_{C^3, \mathcal{B}_{2r}(K_0)} \to 0 \ (k_0 \to \infty)$, for sufficiently large $k_0$, $2d^2\nu^2 |H_{k_0} - H|_{C^3, \mathcal{B}_{2r}(K_0)} \tau < \frac{1}{2}$ holds. Thus for such a $k_0$, $|<\Phi_0>| \le 2d^2\nu^2 |H_{k_0} - H|_{C^3, \mathcal{B}_{2r}(K_0)}$ holds. And from $|<S_0>^{-1}| = \tau_0 < \tau$, $|<\Phi_0(\theta)>| \cdot |<S_0>^{-1}| < \frac{1}{2}$. Hence

$I_n + <S_0>^{-1}<\Phi_0>$ is invertible and therefore $<S_0^{k_0}> = <S_0>[I_n + <S_0>^{-1}<\Phi_0>]$ is invertible. Therefore

$$<S_0^{k_0}>^{-1} = (I + <S_0>^{-1}<\Phi_0>)^{-1}<S_0>^{-1}.$$

And performing some computation, we obtain

$$<S_0^{k_0}>^{-1} = (I_n + <S_0>^{-1}<\Phi_0>)^{-1}(I_n + <S_0>^{-1}<\Phi_0> - <S_0>^{-1}<\Phi_0>)<S_0>^{-1} =$$
$$= <S_0>^{-1} - (I_n + <S_0>^{-1}<\Phi_0>)^{-1}<S_0>^{-1}<\Phi_0><S_0>^{-1}.$$

From the $|(I + <S_0>^{-1}<\Phi_0>)^{-1}| \leq \sum_{i=0}^{\infty}\left|<S_0>^{-1}<\Phi_0>\right|^i < 2$,

$$|<S_0^{k_0}>^{-1}| \leq |<S_0>^{-1}| + 4d^2\nu^2\tau^2 |H_{k_0} - H|_{C^3, \mathcal{B}_{2r}(K_0)}.$$

And $|H_{k_0}|_{C^3, \mathcal{B}_{2r}(K_0)} \leq |H|_{C^3, \mathcal{B}_{2r}(K_0)} + |H_{k_0} - H|_{C^3, \mathcal{B}_{2r}(K_0)}$ holds. On the other hand the sum $\sum_{k=1}^{\infty}|H_k - H_{k-1}|_{C^3, \mathcal{B}_{2r}(K_0)}$ converges, thus for sufficiently large $k_0$, the following inequalities hold

$$2d^2\nu^2 |H_k - H_{k-1}|_{C^3, \mathcal{B}_{2r}(K_0)} \tau < \frac{1}{2}, \quad (k \geq k_0) \tag{8}$$

$$|H_k - H|_{C^3, \mathcal{B}_{2r}(K_0)} < 1, \quad (k \geq k_0) \tag{9}$$

$$4d^2\nu^2\tau^2(|H_{k_0} - H|_{C^3, \mathcal{B}_{2r}(K_0)} + \sum_{k=k_0+1}^{\infty}|H_k - H_{k-1}|_{C^3, \mathcal{B}_{2r}(K_0)}) < 1 \tag{10}$$

Hence the

$$|H_{k_0}|_{C^3, \mathcal{B}_{2r}(K_0)} \leq |H|_{C^3, \mathcal{B}_{2r}(K_0)} + |H_{k_0} - H|_{C^3, \mathcal{B}_{2r}(K_0)} \leq |H|_{C^3, \mathcal{B}_{2r}(K_0)} + 1, |<S_0^{k_0}>^{-1}| < |<S_0>^{-1}| + 1$$

hold and from the definition of $c$,

$$c_{k_0} < c. \tag{11}$$

Performing some computation, we obtain

$$c_{k_0}\gamma^{-4}\delta_0^{-4\sigma} \|e_{k_0}\|_\rho = c_{k_0}\gamma^{-4}\delta_0^{-4\sigma} \|J\nabla H_{k_0}(K_0(\theta)) - \partial_\omega K_0(\theta)\|_\rho =$$
$$\leq c\gamma^{-4}\delta_0^{-4\sigma}(\|J\nabla H_{k_0}(K_0(\theta)) - J\nabla H(K_0(\theta))\|_\rho + \|J\nabla H(K_0(\theta)) - \partial_\omega K_0(\theta)\|_\rho) =$$
$$= c\gamma^{-4}\delta_0^{-4\sigma}(\|J\nabla H_{k_0}(K_0(\theta)) - J\nabla H(K_0(\theta))\|_\rho + \|e_0\|_\rho) \leq$$
$$\leq c\gamma^{-4}\delta_0^{-4\sigma}|H_{k_0} - H|_{C^3, \mathcal{B}_{2r}(K_0)} + c\gamma^{-4}\delta_0^{-4\sigma}\|e_0\|_\rho.$$

From the assumption of theorem 1 : $c\gamma^{-4}\delta_0^{-4\sigma}\|e_0\|_\rho < 1$ and $|H_{k_0} - H|_{C^3, \mathcal{B}_{2r}(K_0)} \to 0 \, (k_0 \to \infty)$, for sufficiently large $k_0$, the following inequality holds

$$c\gamma^{-4}\delta_0^{-4\sigma}|H_{k_0} - H|_{C^3, \mathcal{B}_{2r}(K_0)} + c\gamma^{-4}\delta_0^{-4\sigma}\|e_0\|_\rho < 1.$$

Similarly,

$$c_{k_0}\gamma^{-2}\delta_0^{-2\sigma}\|e_{k_0}\|_\rho \leq c\gamma^{-2}\delta_0^{-2\sigma}|H_{k_0} - H|_{C^3, \mathcal{B}_{2r}(K_0)} + c\gamma^{-2}\delta_0^{-2\sigma}|e_0\|_\rho.$$

And from the assumption of theorem 1 : $c\gamma^{-2}\delta_0^{-2\sigma}|e_0\|_\rho < r$, for sufficiently large $k_0$,

$$c\gamma^{-2}\delta_0^{-2\sigma}|H_{k_0} - H|_{C^3, \mathcal{B}_{2r}(K_0)} + c\gamma^{-2}\delta_0^{-2\sigma}\|e_0\|_\rho < r.$$

Thus

$$c_{k_0}\gamma^{-4}\delta_0^{-4\sigma}\|e_{k_0}\|_\rho < 1 \tag{12}$$

$$c_{k_0}\gamma^{-2}\delta_0^{-2\sigma}\|e_{k_0}\|_\rho < r \tag{13}$$

Therefore for the pair $(H_{k_0}, K_0)$, the assumption of the theorem in section 1 holds. So there exists a solution $K_{k_0}$ of (1) with Hamiltonian $H_{k_0}$ analytic in $U_{\rho/2}$ and degenerate satisfying

$$\| K_{k_0} - K_0 \|_{\rho/2} \leq c\gamma^{-2}\delta_0^{-2\sigma} \| e_0 \|_\rho \leq r. \tag{14}$$

Now we take the number $k_0$ more sufficiently in order that $k_0$ satisfies

$$A(4^{-(k_0-1)})^{l+2\sigma} \leq \| e_0 \|_\rho.$$

Then for any $k \geq k_0$,

$$| H_k - H_{k-1} |_{C^3, \mathcal{B}_{2r}(K_0)} \leq \| e_0 \|_\rho (4^{l+2\sigma})^{-k+1} \tag{15}$$

hold. From now on we consider the subsequence $\{H_k\}_{k=k_0}^\infty$ and for simplicity we denote the sequence $\{H_k\}_{k=k_0}^\infty$ as $\{H_k\}_{k=0}^\infty$.

**Second Step.** For the second step, we will prove the following statements:
For any $k \in \mathbf{N}$, there exists a function $K_k$ analytic in $U_{\rho/2^k}$ which is a solution of (1) with Hamiltonian $H_k$. And it satisfies the inequality

$$\| K_k - K_{k-1} \|_{\rho/4^k} \leq r(\frac{1}{4^{l+\sigma}})^{k-1}.$$

We will prove this, using induction. We define the following notations for function $K_k$ which is obtained by applying the theorem in section 1 $k$-times:

$$\rho_k = \frac{\rho}{2^{k-1}}, \quad \delta_k = \frac{\rho_k}{12}, \quad r_k = r(\frac{1}{4^{l+\sigma}})^{k-1}$$

$$\mathcal{B}_{r_k}(K_k) = \{z \in \mathbf{R}^n ; \inf_{\theta \in U_{\rho_k}} | z - K_k(\theta) | < r_k\}$$

$$e_k(\theta) = J\nabla H_k(K_{k-1}(\theta)) - \partial_\omega K_{k-1}(\theta)$$

$$\mu_k = | H_k |_{C^3, \mathcal{B}_{r_{k-1}}(K_{k-1})}, \quad d_k = \| DK_{k-1} \|_{\rho_{k-1}}, \quad \nu_k = \| N_{k-1} \|_{\rho_{k-1}}, \quad \tau_k = |< S_{k-1}^{H_k} >^{-1}|,$$

and define

$$c_k = \lambda(\mu_k, d_k, \nu_k, \tau_k).$$

To prove the second step, we will prove the following 4 statements, using induction.

A1(k)  $\| K_k - K_0 \|_{\rho_{k+1}} \leq r \sum_{i=0}^{k-1} (\frac{1}{4^{l+\sigma}})^i \leq r \frac{4^{l+\sigma}}{4^{l+\sigma}-1} \leq \frac{4}{3}r$

A2(k)  $c_k \leq c$

A3(k)  $c_k \gamma^{-4} \delta_k^{-4\sigma} \| e_k \|_{\rho_k} < 1$

A4(k)  $c_k \gamma^{-2} \delta_k^{-2\sigma} \| e_k \|_{\rho_k} < r_k$

If A1(k)-A4(k) hold then the assumptions of the theorem in section 1 be satisfied for a pair $(H_k, K_{k-1})$. Then one can obtain $K_k$ from the pair $(H_k, K_{k-1})$ and therefore one obtain a analytic sequence converges to a solution of (1) with Hamiltonian $H$.

First, consider the case of $k=1$. A1(1) is followed by (14) and A2(1), A3(1), A4(1) are respectively followed by (11), (12), (13) in first step.

Assume that A1(j)-A4(j) hold for $j=1, \cdots, k-1$ and let us prove A1(k)-A4(k). For any $j=2, \cdots, k$, we obtain $K_j$, a solution of (1) with Hamiltonian $H_k$, by applying the theorem in section 1 to $K_{j-1}$. Thus inequality (4) implies

$$\| K_k - K_0 \|_{\rho_{k+1}} \leq \| K_k - K_{k-1} \|_{\rho_{k+1}} + \| K_{k-1} - K_0 \|_{\rho_2} \leq$$

$$\leq r_k + r \sum_{i=0}^{k-2} \left(\frac{1}{4^{l+\sigma}}\right)^i \leq r \sum_{i=0}^{k-1} \left(\frac{1}{4^{l+\sigma}}\right)^i$$

This proves A1(k). From the inequality (9),

$$\mu_k = |H_k|_{C^3, \mathcal{B}_{r_{k-1}}(K_{k-1})} \leq |H_k|_{C^3, \mathcal{B}_{2r}(K_0)} \leq |H_0|_{C^3, \mathcal{B}_{2r}(K_0)} + |H_k - H_0|_{C^3, \mathcal{B}_{2r}(K_0)}$$

$$\leq |H_0|_{C^3, \mathcal{B}_{2r}(K_0)} + 1 \leq \mu.$$

And from the statement of theorem in section 1, we obtain

$$d_k = \|DK_{k-1}\|_{\rho_{k-1}} \leq \|DK_{k-2}\|_{\rho_{k-2}} + \gamma^{-2}\delta_{k-1}^{2\sigma-1}2^{-4\sigma} \leq \cdots \leq$$

$$\leq \|DK_0\|_{\rho_0} + \gamma^{-2}\delta_{k-1}^{2\sigma-1}2^{-4\sigma} + \cdots + \gamma^{-2}\delta_0^{2\sigma-1}2^{-4\sigma} \leq$$

$$\leq \|DK_0\|_{\rho_0} + \sum_{i=0}^{\infty}\gamma^{-2}\delta_i^{2\sigma-1}2^{-4\sigma} \leq \|DK_0\|_{\rho_0} + \gamma^{-2}\delta_0^{2\sigma-1}2^{-4\sigma}\sum_{i=0}^{\infty}(1/2^i)^{2\sigma-1} \leq$$

$$\leq \|DK_0\|_{\rho_0} + \gamma^{-2}\delta_0^{2\sigma-1}2^{-4\sigma}\frac{1}{1-2^{-2\sigma+1}} \leq \|DK_0\|_{\rho_0} + \gamma^{-2}\delta_0^{2\sigma-1}\frac{1}{2^{4\sigma}-2^{2\sigma+1}}$$

$$\leq d_0 + \beta = d.$$

Similarly, we get

$$\nu_k = \|N_{k-1}\|_{\rho_{k-1}} \leq \nu_0 + \beta = \nu.$$

Now let us prove the following inequalities by using induction:

$$\tau_k = |<S_{k-1}^{H_k}>^{-1}| \leq \tau_0 + \beta + 1 = \tau,$$

$$|<S_{k-1}^{H_{k-1}}>^{-1}| \leq \tau_0 + \beta + 1 = \tau.$$

In the case of $k=1$ is trivial. Assume that they stand for $i = 0, \cdots, k-1$. Performing similar computation in step 1, we obtain

$$S_{k-1}^{H_k}(\theta) = N_{k-1}(\theta)DK_{k-1}(\theta)^T[A_k(\theta)J - JA_k(\theta)]DK_{k-1}(\theta)N_{k-1}(\theta) =$$

$$= N_{k-1}(\theta)DK_{k-1}(\theta)^T[(A_{k-1}(\theta)J - JA_{k-1}(\theta)) + (A_k(\theta) - A_{k-1}(\theta))J - J(A_k(\theta) - A_{k-1}(\theta))],$$

$$DK_{k-1}(\theta)N_{k-1}(\theta) = S_{k-1}^{H_{k-1}}(\theta) + N_{k-1}(\theta)DK_{k-1}(\theta)^T[(A_k(\theta) - A_{k-1}(\theta))J - J(A_k(\theta) - A_{k-1}(\theta))],$$

$$DK_{k-1}(\theta)N_{k-1}(\theta) = S_{k-1}^{H_{k-1}}(\theta) + \Phi_{k-1}(\theta)$$

and

$$\|\Phi_{k-1}(\theta)\|_{\rho_{k-1}} \leq 2d_k^2\nu_k^2|H_k - H_{k-1}|_{C^3, \mathcal{B}_{r_{k-1}}(K_{k-1})} \leq 2d^2\nu^2|H_k - H_{k-1}|_{C^3, \mathcal{B}_{r_{k-1}}(K_{k-1})}.$$

From (8), the inequality

$$2d_k^2\nu_k^2|H_k - H_{k-1}|_{C^3, \mathcal{B}_{2r}(K_0)}|<S_{k-1}^{H_{k-1}}>^{-1}| < 2d^2\nu^2|H_k - H_{k-1}|_{C^3, \mathcal{B}_{r_{k-1}}(K_{k-1})}\tau \leq \frac{1}{2}$$

follows. Thus

$$|<\Phi_{k-1}>|\cdot|<S_{k-1}^{H_{k-1}}>^{-1}| \leq \frac{1}{2}.$$

Hence $I_n + <S_{k-1}^{H_{k-1}}>^{-1}<\Phi_{k-1}>$ is invertible and therefore

$$<S_{k-1}^{H_k}> = <S_{k-1}^{H_{k-1}}>(I_n + <S_{k-1}^{H_{k-1}}>^{-1}<\Phi_{k-1}>)$$

is invertible, i.e.

$$<S_{k-1}^{H_k}>^{-1} = (I_n + <S_{k-1}^{H_{k-1}}>^{-1}<\Phi_{k-1}>)^{-1}<S_{k-1}^{H_{k-1}}>^{-1}.$$

And performing some computation, we obtain

$$<S_{k-1}^{H_k}>^{-1} =$$

$$= (I_n + <S_{k-1}^{H_{k-1}}>^{-1}<\Phi_{k-1}>)^{-1}(I_n + <S_{k-1}^{H_{k-1}}>^{-1}<\Phi_{k-1}> - <S_{k-1}^{H_{k-1}}>^{-1}<\Phi_{k-1}>)<S_{k-1}^{H_{k-1}}>^{-1} =$$

$$= <S_{k-1}^{H_{k-1}}>^{-1} + (I_n + <S_{k-1}^{H_{k-1}}>^{-1}<\Phi_{k-1}>)^{-1}<S_{k-1}^{H_{k-1}}>^{-1}<\Phi_{k-1}><S_{k-1}^{H_{k-1}}>^{-1}.$$

The inequality

$$|(I+<S_{k-1}^{H_{k-1}}>^{-1}<\Phi_{k-1}>)^{-1}|\leq\sum_{i=0}^{\infty}|<S_{k-1}^{H_{k-1}}>^{-1}<\Phi_{k-1}>|^{i}<2$$

Holds and thus we have

$$|<S_{k-1}^{H_k}>^{-1}|\leq|<S_{k-1}^{H_{k-1}}>^{-1}|+4d^2\nu^2\tau^2|H_k-H_{k-1}|_{C^3,\mathcal{B}_{r_{k-1}}(K_{k-1})}.$$

And from the theorem in section 1, the inequality

$$|<S_{k-1}^{H_{k-1}}>^{-1}|<|<S_{k-2}^{H_{k-1}}>^{-1}|+\gamma^{-2}\delta_{k-1}^{2\sigma-1}2^{-4\sigma}$$

holds. Repeating the above inequality, we obtain

$$|<S_{k-1}^{H_k}>^{-1}|\leq|<S_{k-2}^{H_{k-1}}>^{-1}|+4d^2\nu^2\tau^2|H_{k_0}-H|_{C^3,\mathcal{B}_{2r}(K_0)}+\gamma^{-2}\delta_{k-1}^{2\sigma-1}2^{-4\sigma}\leq\cdots\leq$$

$$\leq|<S_0^{H_0}>^{-1}|+\sum_{j=1}^{k}4d^2\nu^2\tau^2|H_j-H_{j-1}|_{C^3,\mathcal{B}_{2r}(K_0)}+\sum_{j=1}^{k}\gamma^{-2}\delta_{j-1}^{2\sigma-1}2^{-4\sigma}\leq$$

$$\leq|<S_0^{H_0}>^{-1}|+\sum_{j=1}^{\infty}4d^2\nu^2\tau^2|H_j-H_{j-1}|_{C^3,\mathcal{B}_{2r}(K_0)}+\sum_{j=1}^{\infty}\gamma^{-2}\delta_{j-1}^{2\sigma-1}2^{-4\sigma}\leq$$

$$\leq|<S_0^{H_0}>^{-1}|+1+\beta.$$

(here, we use (10)). The estimation

$$|<S_{k-1}^{H_{k-1}}>^{-1}|\leq|<S_0^{H_0}>^{-1}|+1+\beta$$

has been already proved in the process of the proof. Therefore for any $k\in\mathbf{N}$, the inequality

$$\tau_k=|(avg\{S_{k-1}^{H_k}\}_\theta)^{-1}|\leq\tau$$

holds and from the definition of $c$, we have

$$c_k\leq c.$$

This implies A2(k).

Next let us prove A3(k). Performing some computation, we obtain

$$c_k\gamma^{-4}\delta_k^{-4\sigma}\|e_k\|_{\rho_k}=c\gamma^{-4}(\delta_0/2^{k-1})^{-4\sigma}\|J\nabla H_k(K_{k-1}(\theta))-\partial_\omega K_{k-1}(\theta)\|_{\rho_k}=$$

$$\leq c\gamma^{-4}(\delta_0/2^{k-1})^{-4\sigma}(\|J\nabla H_k(K_{k-1}(\theta))-J\nabla H_{k-1}(K_{k-1}(\theta))\|_{\rho_k}+$$

$$+\|J\nabla H_{k-1}(K_{k-1}(\theta))-\partial_\omega K_{k-1}(\theta)\|_{\rho_k})=$$

$$=c\gamma^{-4}\delta_0^{-4\sigma}(2^{k-1})^{4\sigma}\|J\nabla H_k(K_{k-1}(\theta))-J\nabla H_{k-1}(K_{k-1}(\theta))\|_{\rho_k}\leq$$

$$\leq c\gamma^{-4}\delta_0^{-4\sigma}|H_k-H_{k-1}|_{C^3,\mathcal{B}_{r_{k-1}}(K_{k-1})}(2^{k-1})^{4\sigma}\leq$$

$$\leq c\gamma^{-4}\delta_0^{-4\sigma}\|e_0\|_{\rho_0}(4^{l+2\sigma})^{-k+1}(2^{k-1})^{4\sigma}\leq$$

$$\leq(4^{l+2\sigma})^{-k+1}(4^{k-1})^{2\sigma}\leq(\frac{1}{4^l})^{k-1}\leq1.$$

(here, the inequality $|H_k-H_{k-1}|_{C^3,\mathcal{B}_{r_{k-1}}(K_{k-1})}\leq\|e_0\|_{\rho_0}(4^{l-3})^{-k+1}$ is followed by (14).) Therefore A3(k) holds. Similarly,

$$c_k\gamma^{-2}\delta_k^{-2\sigma}\|e_k\|_{\rho_k}=c\gamma^{-2}(\delta_0/2^{k-1})^{-2\sigma}\|J\nabla H_k(K_{k-1}(\theta))-\partial_\omega K_{k-1}(\theta)\|_{\rho_k}=$$

$$\leq c\gamma^{-2}\left(\frac{\delta_0}{2^{k-1}}\right)^{-2\sigma}(\|J\nabla H_k(K_{k-1}(\theta))-J\nabla H_{k-1}(K_{k-1}(\theta))\|_{\rho_k}+\|J\nabla H_{k-1}(K_{k-1}(\theta))-\partial_\omega K_{k-1}(\theta)\|_{\rho_k})$$

$$=c\gamma^{-2}\delta_0^{-2\sigma}(2^{k-1})^{2\sigma}\|J\nabla H_k(K_{k-1}(\theta))-J\nabla H_{k-1}(K_{k-1}(\theta))\|_{\rho_k}\leq$$

$$\leq c\gamma^{-2}\delta_0^{-2\sigma}|H_k-H_{k-1}|_{C^3,\mathcal{B}_{r_{k-1}}(K_{k-1})}(2^{k-1})^{2\sigma}\leq$$

$$\leq c\gamma^{-2}\delta_0^{-2\sigma}\|e_0\|_{\rho_0}(4^{l+2\sigma})^{-k+1}(2^{k-1})^{2\sigma}\leq$$

$$\leq r(4^{l+2\sigma})^{-k+1}(4^{k-1})^\sigma\leq r(1/4^{l+\sigma})^{k-1}=r_k,$$

and this implies A4(k). Hence we have proved A1(k)-A4(k) for all $k \in \mathbf{N}$.

$K_k$ which is obtained by applying the theorem in section 1 $k$-times satisfy the following inequality:

$$\| K_k - K_{k-1} \|_{\rho/4^k} \leq \| K_k - K_{k-1} \|_{\rho_k} \leq r_k = r\left(\frac{1}{4^{l+\sigma}}\right)^{k-1}.$$

Therefore from lemma 4, the sequence $\{K_k\}$ converges to $K_\infty \in C^1(\mathbf{T}^n)$.

Because for any $k \in \mathbf{N}$, $K_k$ satisfy (1) with Hamiltonian $H_k$ and $H_k$ converges to $K_\infty$ in $\mathbf{T}^n$, therefore $K_\infty$ is a solution of (1) with $H$.

**Remark 3**. In theorem 1, we assumed standard symplectic forms like Rüssmann, Pöschel, Llave; the reason is that once it is done, the general case will be done through some standard but complicated calculation. [Llave et al 2008] is close to our result. They considered finitely differentiable symplectic **maps** without Action-Angle Variables. In this article, we considered finitely differentiable symplectic **vector fields** without Action-Angle Variables. [Llave et al 2008]'s result can be applied to Periodic vector fields but can't be applied to nonperiodic vector fields. As Llave mentioned in his letter to one of authors, smoothing periodic vector fields is easier.

**Acknowledgement**: Authors would like to thank the anonymous reviewers and de la Llave, Rafael for their useful advice and encouragement.

# References

[Arnold 1963] Arnold V.I.: Proof of a Theorem of A.N.Kolmogorov on the Invariance of Quasi-Periodic Motions under Small Perturbations of the Hamiltonian, Russian Mathematicla Surveys 18(8), (1963) 9-36.

[Arnold, 1968] Arnold V.I., Avez A.: Ergodic Problems of Classical Mechanics, Benjamin, 1968.

[Arnold 1989] Arnold V.I.: Mathematicla Methods of Classical Mechanics, Springer, 1989.

[Broer, 2008] Broer H. W., Sevryuk M. B.; KAM Theory: quasi-periodicity in dynamical systems, 2008.

[Casartelli 1983] Casartelli M.: Relaxation Times and Randomness for a Nonlinear Classical System, Lecture Notes in Physics 179, Springer, (1983), 252-253.

[Celletti, 2006] Celletti A., Chierchia L.: KAM Stability for a three-body problem of the Solar system, Z. angew. Math. Phys. 57 (2006) 33–41.

[Chierchia 2007] Chierchia: KAM Theory, Universita "Roma Tre", (2007), 1-51.

[Fèjoz 2004] Fèjoz J.: Démonstration du « théor ème d'Arnold » sur la stabilité du système planétaire (d'après M. Herman), Ergod. Th. & Dynam. Sys. (2004),24, 1-62.

[Gidea, 2009] Gidea M., Meiss J. D., Ugarcovicic I., Weiss H.: Applications of KAM Theory to Population Dynamics, Journal of Biological Dynamics, (2009), 1–23.

[Gonz 2008] Gonzalez-Enriquez, A. and Vano, J., An estimate of smoothing and composition with applications to conjugation problems, J. Dynam. Differential Equations, VOLUME 20, 2008, NUMBER 1, PAGES 239~270, DOI = {10.1007/s10884-006-9060-z}, URL http://dx.doi.org/10.1007/s10884-006-9060-z

[Hairer, 2006] Hairer E., Lubich C., Wanner G.: Geometric Numerical Integration - Structure-Preserving Algorithms for Ordinary Differential Equations, Springer, 2006.

[Haro, 2004] Haro A and de la Llave R: A parameterization method for the computation of whiskers in quasi periodic maps: numerical implementation Preprint mp arc 04-350, 2004.

[Helleman, 1985] Helleman R., Kheifets S. A.: Nonlinear Dynamics Aspects of Modern Storage Rings, Lecture Notes in Physics 247, Springer, (1985), 64-76.

[Herman 1986] Herman M.R.: Sur les courbes invariantes par les difféomorphismes de l'anneau, Vol. 1 and 2, Astérisque 103–104 (1983), i and 1–221; 144 (1986), 1–248.

[Llave, 2005] Llave R. de la, González A., Jorba, À., Villanueva J.: KAM theory without action-angle variables, Nonlinearity 18 (2005), 855–895.

[Llave 2008] Gonzalez-Enriquez, A. and de la Llave, R., Analytic smoothing of geometric maps with applications to KAM theory, J. Differential Equations, VOLUME 245, 2008, NUMBER 5, PAGES 1243~1298, DOI = {10.1016/j.jde.2008.05.009}, URL http://dx.doi.org/10.1016/j.jde.2008.05.009,

[Llave 2001] de la Llave, Rafael, A tutorial on KAM theory, Smooth ergodic theory and its applications (Seattle, WA, 1999), Proc. Sympos. Pure Math., VOLUME 69, PAGES 175~292, Amer. Math. Soc., Providence, RI, 2001,

[Locatelli, 2007] Locatelli U., Giorgilli A.: Invariant tori in the Sun-Jupiter-Saturn System, AIM science.org Discrete and Continuous Dynamical Systems-Sires B, Vol. 7, No 2, March (2007), 377-398.

[Lorentz 1986] Lorentz G.G.: Bernstein Polynomials 2ed, Chelsea Publishing Company, 1986.

[Moser 1961] Moser J.K.: A new technique for the construction of solutions of nonlinear differential equations. Proc. Nat. Acad. Sci. U.S.A. 47 (1961), 1824–1831.

[Moser 1962] Moser J.K.: On invariant curves of area-preserving mappings of an annulus, Nach. Akad. Wiss. Göttingen, Math. Phys. Kl. II 1 (1962), 1-20.

[Moser 1966] Moser J.K.: A rapidly convergent iteration method and non-linear differential equations, I, Annali della Scuola Norm. Super, de Pisa ser. III, 20, (1966), 265-315; II, (1966), 499-535.

[Rüssmann 1970] Rüssmann H.: Kleine Nenner I: Über invariante Kurven differenzierbarer Abbildungen eines Kreisringes, Nachr. Akad.Wiss. Göttingen,Math.-Phys. Kl. II 5 (1970), 67–105.

[Salamon2004] Salamon, Dietmar A., The Kolmogorov-Arnold-Moser Theorem, Math. Phys. Electron. J., VOLUME 10, 2004, 37 pp.(electronic)

[Takens 1971] Takens F.: A $C^1$ counterexample to Moser's twist theorem, Indag. Math. 33 (1971), 378–386.

[Zehnder 75] Zehnder, E., Generalized implicit function theorems with applications to some small divisor problems. I, Comm. Pure Appl. Math., VOLUME 28,1975, PAGES 91~140,